\documentclass{INTERSPEECH2023}
\usepackage{multirow}
\usepackage{cite}
\usepackage{setspace}
\usepackage[dvipsnames]{xcolor}
\title{Hyper-parameter Adaptation of Conformer ASR Systems\\ for Elderly and Dysarthric Speech Recognition}
\interspeechcameraready
\ninept
\name{Tianzi Wang$^{1}$, Shoukang Hu$^{2}$, Jiajun Deng$^{1}$, Zengrui Jin$^{1}$, Mengzhe Geng$^{1}$, Yi Wang$^{1}$,\\{Helen Meng$^{1}$, Xunying Liu$^{1}$}}
\address{
$^{1}$The Chinese University of Hong Kong, China\\
$^{2}$Nanyang Technological University, Singapore 
}
\email{$^{1}$\{twang,jjdeng,zrjin,mzgeng,ywang,hmmeng,xyliu\}@se.cuhk.edu.hk,\\$^{2}$shoukang.hu@ntu.edu.sg}
\begin{document}
\bstctlcite{IEEEexample:BSTcontrol}

\maketitle
 
\begin{abstract}
Automatic recognition of disordered and elderly speech remains highly challenging tasks to date due to data scarcity. Parameter fine-tuning is often used to exploit the large quantities of non-aged and healthy speech pre-trained models, while neural architecture hyper-parameters are set using expert knowledge and remain unchanged. This paper investigates hyper-parameter adaptation for Conformer ASR systems that are pre-trained on the Librispeech corpus before being domain adapted to the DementiaBank elderly and UASpeech dysarthric speech datasets. Experimental results suggest that hyper-parameter adaptation produced word error rate (WER) reductions of 0.45\% and 0.67\% over parameter-only fine-tuning on DBank and UASpeech tasks respectively. An intuitive correlation is found between the performance improvements by hyper-parameter domain adaptation and the relative utterance length ratio between the source and target domain data.
\end{abstract}

% Automatic recognition of disordered and elderly speech remain highly challenging tasks to date due to difficulty in collecting such data. In order to exploit large quantities of non-aged and healthy adult speech pre-trained ASR systems, parameter fine-tuning based cross-domain adaptation is often used. The neural architecture hyper-parameters that are often set using expert and domain knowledge remain unchanged. This paper investigates architectural hyper-parameter adaptation for Conformer ASR systems that are pre-trained on the 960-hour Librispeech corpus before being domain adapted to the 16-hour DementiaBank Pitt elderly and 31-hour UASpeech dysarthric speech datasets. Hyper-paramter adaptation is reformulated as a cross domain differentiable neural architecture search (DARTS) task. Experimental results suggest that architecture adaptation produced significant word error rate (WER) reductions on the DementiaBank and UASpeech test data, where a large mismatch in terms of both utterance lengths and speech contents against the  data is found. 

\noindent\textbf{Index Terms}: Dysarthric Speech, Elderly Speech,  Conformer, Domain Adaptation, Hyper-parameter Adaptation

\section{Introduction}

Despite the rapid progress of automatic speech recognition (ASR) technologies targeting normal speech, accurate recognition of elderly and dysarthric speech remains a challenging task \cite{christensen2013combining,yu2018development,xiong2020source,liu2020exploiting,ye2021development,geng2022spectro,deng2021bayesian,geng2022speaker}. %% Aging presents enormous challenges to health care worldwide. 
Neurocognitive disorders, e.g. Alzheimer’s disease (AD), are often found among older adults \cite{alzheimer20192019} experiencing speech impairments \cite{fraser2016linguistic,konig2018fully}. ASR technologies trailed for their needs can improve their quality of life and social inclusion. 

Elderly and dysarthric speech exhibit a wide spectrum of challenges for current deep neural networks (DNNs) based ASR technologies that predominantly target normal speech. First, a large mismatch between such data and non-aged, healthy adult voices is often observed. Such difference manifests itself across many fronts including articulatory imprecision, decreased volume and clarity, changes in pitch, increased dysfluencies, and slower speaking rate \cite{hixon1964restricted,kent2000dysarthrias}. Second, the co-occurring disabilities and mobility issues among elderly speakers lead to the difficulty in collecting large quantities of such data that are essential for current data-intensive ASR system development. 

A widely adopted solution to address above issues for dysarthric and elderly speech recognition is to use model-based domain adaptation approaches \cite{xiong2020source,green2021automatic,geng2020investigation,christensen2013combining,wang2021improved,liu2021recent}. 
%% which allows widely available normal speech to be leveraged to facilitate cross domain knowledge. 
Current practice of domain adaptation of E2E ASR systems mainly perform parameter fine-tuning, while the underlying architecture hyper-parameters 
%% which are often set using expert knowledge and empirical evaluation for specific task domains, 
remain unchanged during domain adaptation.
However, previous studies suggest that the optimal architectural hyper-parameters are heavily domain specific. For example, in \cite{deng2021bayesian} demonstrated the optimal settings of hidden layer context offsets in hybrid TDNN acoustic models, which are used to encode hidden layer level temporal context spans, vary significantly between TDNN systems optimized for fluent, normal speech of longer utterances, and those constructed on disfluent, impaired speech utterances of single word commands or short phrases. A similar study that performed architecture adaptation was conducted on CTC-based CNN ASR systems \cite{chen2020darts} for multilingual speech recognition, revealing that optimal convolutional module hyper-parameters, e.g. the convolution kernal size, vary substantially between languages. In contrast, the hyper-parameters domain adaptation of state-of-the-art end-to-end ASR systems represented by, for example, those based on Conformer models \cite{gulati2020conformer,zhang2020pushing,guo2021recent,hsu2021robust,yao2021wenet,zhang2020unified,zhang2022bigssl,chan2021speechstew}, remains unvisited for dysarthric and elderly speech recognition. 

To this end, this paper presents the study on cross-domain hyper-parameter adaptation from Librispeech \cite{panayotov2015librispeech} pre-trained Conformer ASR systems to two atypical speech recognition tasks: a) the 16-hour DementiaBank Pitt elderly speech corpus (DBank) \cite{becker1994natural}; and b) the 31-hour UASpeech dysarthric speech \cite{kim2008dysarthric} dataset. In this work, the hyper-parameter domain adaptation problem is transformed into a cross domain differentiable neural architecture search (DARTS) \cite{liu2018darts} task. A DARTS super-network is constructed to contain all possible candidate structures associated with varying Conformer encoder and decoder hyper-parameters settings: a) the dimensionality of macron-feedforward and feedforward layers; b) the number and the dimensionality of attention heads; and c) the kernel size of convolution modules. Thus super-network is initially pre-trained on the 960-hour Librispeech corpus with auto-configured hyper-parameter weights, before being domain adapted to the DBank or UASpeech data to extract the optimal target domain specific hyper-parameters. Conformer systems configured using the resulting hyper-parameters setting are then parameter wise pre-trained on the Librispeech data before being cross domain fine-tuned to the DBank or UASpeech datasets respectively. 

Experimental results suggest that hyper-parameter domain adaptation produced consistent 
% and statistically significant\footnote{Matched pairs sentence-segment word error (MAPSSWE) based statistical signiﬁcance tests \cite{gillick1989some} at signiﬁcance level $\alpha=0.05$.} 
word error rate (WER) reductions of 0.45\% and 0.67\% absolute (1.81\% and 2.37\% relative) for the DBank and UASpeech tasks respectively over the conventional parameter fine-tuning only domain adaptation Conformer systems. Such WER reduction from Conformer hyper-parameter adaptation is found to be intuitively correlated with the average utterance length ratio between the source and target domain data, e.g. Librispeech vs. DBank (12.3s:3.4s) data while Librispeech vs. UASpeech (12.3s:1.1s). 
%% on the target utterances with short temporal context than the other utterances. since the average utterance length of pretraining Librispeech (12.3s) is much longer than the target Dbank (3.4s) and UASpeech (1.1s), 
This correlation is further validated by evaluating the WER reductions on the subsets of DBank and UASPeech test data of shorter and longer utterance lengths. 
%% which are separated by median utterance length from the original test sets of DBank and UASpeech. 
On the shorter segments based subsets of DBank and UASpeech test data, hyper-parameter adaptation produces larger WER reductions of 1.2\% and 0.9\% respectively over parameter-only fine-tuning. In contrast, smaller WER reductions of 0.3\%-0.4\% were obtained on the longer segments based subsets. Such correlation is also consistent with the corresponding larger or smaller changes of Conformer hyper-parameters: a) the number of attention heads that affect longer range contexts; and b) the kernel sizes of the convolution modules designed to control the span of local contexts. 
%% hyper-parameter adaptation is found less important on longer utterances, leading to only 0.3\%-0.4\% WER reduction. 

To the best of our knowledge, this paper presents the first investigation on the hyper-parameter adaptation in Conformer systems for elderly and dysarthric speech recognition. In contrast, the majority of previous research on domain adaptation for the same tasks has been focused on direct parameter fine-tuning \cite{pan2021using,lin2020staged,shor2019personalizing,xiong2020source}. Limited previous researches on ASR architecture adaptation were conducted on hybrid TDNN models \cite{deng2021bayesian}, or CTC-based CNN multilingual ASR \cite{chen2020darts}. For Conformer ASR systems, only in-domain data based neural architectural search were studied for elderly \cite{wang2022conformer} and Mandarin \cite{shi2021darts} speech recognition, while the cross domain architecture adaptation problem was not considered. 
\begin{figure*}[htb]
    \label{fig:pipeline}
    \centering
    \includegraphics[width=1.0\textwidth]{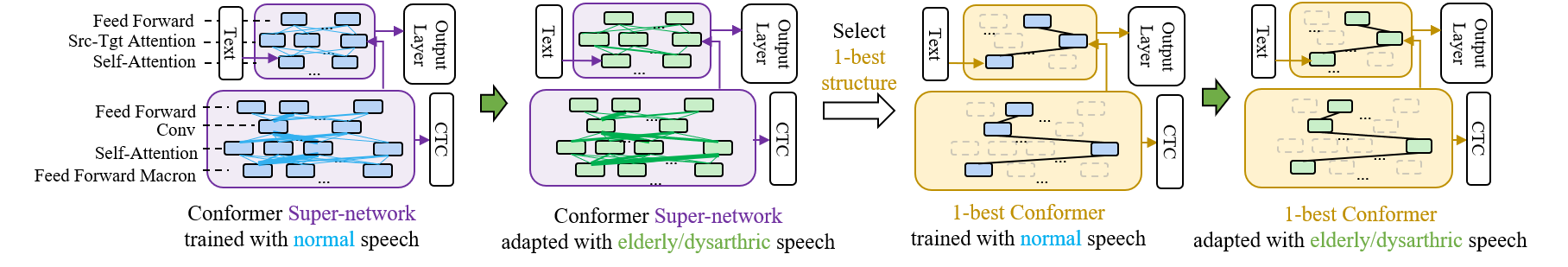}
    \vspace{-0.8cm}
    \caption{\rm \setstretch{0.9}\fontsize{7.5bp}{7.5bp}\textnormal{The pipeline of parameter and hyper-parameter domain adaptation of Conformer ASR systems for elderly/dysarthric speech recognition, including adaptation of hyper-parameters inside the super-network model decoupled with standard network parameters (left, blue) from normal speech to elderly/dysarthric speech (green), and parameter adaptation of 1-best domain adapted hyper-parameter based Conformer (right, yellow) with the same source and target speech.}}\label{fig:pipeline}
    \vspace{-0.7cm}
\end{figure*}
\vspace{-0.2cm}
\section{Baseline and Parameter Adaptation}
\subsection{Conformer Baseline Architecture}
This section reviews the Conformer architecture and presents the baseline parameter fine-tuning setup on two atypical ASR tasks: a) the 16-hour DBank elderly speech corpus\cite{becker1994natural}; and b) the 31-hour UASpeech dysarthric speech dataset \cite{kim2008dysarthric}. \\
\textbf{Hybrid CTC/attention based Conformer} A novel structure named Conformer was proposed in \cite{gulati2020conformer} and achieved state-of-the-art results on LibriSpeech ASR tasks. 
%Inside the Conformer encoder block, macron-feedforward layers, a multi-head self-attention layer, a convolution layer, and a feedforward layer are sequentially connected to learn the global context and local correlations synchronously. Each Transformer decoder block consists of a multi-head source-target attention layer, multi-head self-attention layer, and a feedforward layer.
In this work, we built the sequence trained end-to-end Conformer ASR following the ESPnet \cite{watanabe2018espnet} hybrid CTC/attention encoder-decoder structure. The encoder contains 2 Convolution blocks to downsample the 80-dimension Mel-scale filter banks and 3-dimension pitch inputs, followed by stacked Conformer blocks. An interpolated CTC+AED cost function (3:7 weighting) was computed over the output vocabulary. The hyper-parameters settings of baseline Conformer system\footnote{12 Conformer encoder blocks + 6 Transformer decoder blocks, feed-forward layer dim = 2048, attention heads = 4, dim of attention heads = 256, convolution kernel size=31} for both DBank and UASpeech are determined following the ESPnet Switchboard recipe.
\vspace{-0.2cm}
\subsection{Parameter Adaptation}
In order to exploit large quantities of out-of-domain, non-aged and healthy adult speech pre-trained Conformer systems, parameter fine-tuning based cross-domain adaptation was considered. A 960-hour Librispeech corpus trained Conformer system was parameter fine-tuned to the DBank or UASpeech data after speed perturbation based data augmentation \cite{geng2020investigation}.
\vspace{-0.2cm}
\section{Hyper-parameter Adaptation}
\subsection{Differentiable Architecture Search}
To automatically learn the suitable Conformer hyper-parameter settings for the target atypical speech domain, DARTS \cite{liu2018darts} was used to optimize four groups of hyper-parameters inside each Conformer encoder block: a) the feedforward layer dimensionality (FD); b) the number of attention heads (AH); c) the dimensionality of attention head (ADIM); d) the convolution kernel size (CK), as well as three groups of hyper-parameters inside each Transformer decoder block (FD, AH and ADIM).
The estimation of standard network parameters inside the super-network model is decoupled from that of the architecture parameters \cite{hu2022neural, guo2020single}. This leads to a pipelined approach allowing the architectural weights to be learned on separate held-out data. The optimal architecture with the largest weight is selected. 
%The general for\begin{equation}
%\vspace{-0.1cm}
%     x^l=\sum_{i=1}^{N^l}\lambda_{i}^{l}c_i^l(x^{l-1})=\sum_{i=1}^{N^l}\frac{\text{exp}(\alpha_i^{l})}{\sum_{j=1}^{N^l}\text{exp}(\alpha_j^{l})}c_i^l(x^{l-1})
%\end{equation}m of DARTS architecture selection is as follows: 

%where $c_i^l$ and $\lambda_i^l$ are the i-th candidate architecture choice of l-th layer and its corresponding weight respectively. $\lambda_i^l$ is modelled by a Softmax function over a vector $\alpha^l$, whose dimensionality equals to the total number of candidate architectures, $N^{l}$. 
\noindent\textbf{Gumbel-Softmax DARTS}:
In traditional DARTS methods, when similar architecture weights are obtained using a flattened Softmax function, the confusion over different candidate systems increases and search errors may occur. To this end, a Gumbel-Softmax distribution \cite{xie2018snas, hu2022neural, hu2020dsnas} is used to sharpen the architecture weights to produce approximately a one-hot vector. The architecture weights are computed as,
\begin{equation}
\setlength{\abovedisplayskip}{0.5pt}
	\setlength{\belowdisplayskip}{0.5pt}
     \lambda_{i}^{l}=\frac{{\rm exp}(\log(\alpha_i^{l}+G_{i}^l)/T)}{\sum_{j=1}^{N^l}{\rm exp}(\log(\alpha_j^{l}+G_j^{l})/T)}
     \label{eqn:gumble}
\end{equation}
where $G_{i}^{l}=-\log(-\log(U_i^l))$ is the Gumbel variables and $U_i^l$ is a uniform random variable. As the temperature factor $T$ decreases to zero, Eqn. (\ref{eqn:gumble})  approaches a categorical distribution.

\noindent\textbf{Penalized DARTS}:
In order to avoid over-parameterization
during architecture search, a penalty loss incorporating the number of parameters for each candidate choice was jointly optimized with the original Conformer training loss function:
\begin{equation}
\setlength{\abovedisplayskip}{0.5pt}
	\setlength{\belowdisplayskip}{0.5pt}
     \mathcal{L}=\mathcal{L}_{Conformer} + \eta\sum\nolimits_{i, l}\alpha_{i}^{l}P_{i}^{l}
\end{equation}
where $P_{i}^{l}$ is the number of parameters of the i-th candidate architecture at the l-th layer, and $\eta$ is the penalty scaling factor empirically adjusted for performance vs. complexity trade-off. 
\vspace{-0.2cm}
\subsection{Hyper-parameter Adaptation}
Hyper-parameter adaptation of Conformer systems is performed in two stages: \textbf{1)} A super-network shown in Fig. \ref{fig:pipeline} (left, purple) that contains all possible candidate structures associated with varying hyper-parameter settings of each encoder block (FD, AH, ADIM, CK) and decoder block (FD, AH, ADIM) using the source domain Librispeech data alone, before being adapted to the target elderly or dysarthric speech domain. In this process, the large number of standard Conformer super-network parameters, often in tens of millions, are inherited and adapted to the limited target domain data, while the comparatively much smaller number of hyper-parameter selection weights,
% as shown in Tab. \ref{tab:dbank},
are fine-tuned during domain adaptation.
\textbf{2)} The differentiable architecture search performed over the resulting domain adapted Conformer super-network will then produce the 1-best hyper-parameter settings with the largest weights. A Conformer system that features the above 1-best hyper-parameter configurations using the source domain training data is then constructed. This is followed by the parameter fine-tuning based domain adaptation as described in Section 2.2. The standard model parameters of this source domain Conformer system are then further adapted via fine-tuning to the target domain speech to produce the full hyper-parameter plus parameter adapted system (Fig. \ref{fig:pipeline}, right, yellow). 
\vspace{-0.2cm}
\section{Experiments}
The proposed auto-configurable neural architectural and parametric domain adaptation approach was investigated from the Librispeech data with 960-hour audiobooks speech collected from 2338 speakers to two tasks of Conformer systems: a) the 16-hour DementiaBank (DBank) elderly speech corpus; and b) the 31-hour UASpeech dysarthric speech dataset. \\
\noindent\textbf{DementiaBank elderly speech} The DementiaBank (DBank) elderly speech database \cite{becker1994natural} consists of 16-hours of a training set (29682 utterances) recorded over interviews between the 292 elderly participants (Par) and the clinical investigators (Inv) after silence stripping \cite{ye2021development}, and is expanded to 59-hours when speed perturbation was performed \cite{geng2020investigation}. The development and evaluation sets contain 2.5-hours and 0.6-hours of audio respectively. The source-target domain utterance length ratio between Librispeech and DBank is 12.3s:3.4s on average. A 4-gram character language model built with the DBank transcripts was used. \\
%The DBank database \cite{becker1994 natural} consists of 16 hours of training set a (29682 utterances) recorded over interviews between the 292 elderly participants and the clinical investigators after silence stripping \cite{ye2021development}, and is expanded to 59-hours when speed perturbation was performed \cite{geng2022investigation}. The development and evaluation sets contain 2.5-hours and 0.6-hours of audio respectively. The source-target domain utterance length ratio between Librispeech and DBank is 5.3:3.4 on average. A 4-gram character language model described in \cite{ye2021development} was used. \\
\textbf{UASpeech dysarthric speech} \cite{kim2008dysarthric,christensen2012comparative} is an isolated word recognition task (including 155 common words and 300 uncommon words) that is collected from 16 dysarthric and 13 control speakers using multiple microphone channels \cite{christensen2012comparative}. The training data covers all 29 speakers contains all 155 common words and two third of the uncommon words (31 hours after silence stripping), while the evaluation data was collected from only the 16 dysarthric speakers and cover all 155 common words and the remaining one-third of the uncommon words (9-hours after silence stripping). As E2E ASR systems are sensitive to the training data coverage, B2 data of the $13$ control speakers are also used in Conformer training and cross domain adaptation. This produces a $40$ hour unaugmented ($122392$ utt.) and a $190$-hour augmented training set ($538292$ utt.) after applying speaker independent and dependent speed perturbation \cite{geng2020investigation}.
The source-target domain utterance length ratio between Librispeech and UASpeech is 12.3s:1.1s on average. A word grammar language model was used in decoding \cite{christensen2012comparative}.\\
%% The UASpeech disordered speech \cite{kim2008dysarthric} is an isolated word recognition task (including 155 common words and 300 uncommon words) that is collected from 16 dysarthric and 13 control speakers. \textcolor{red}{Unlike previous Hybrid ASR systems for UASpeech \cite{xiong2019phonetic,liu2019use,christensen2012comparative}, in which the training data covers all 29 speakers contains all 155 common words and two third of the uncommon words, E2E ASR systems were found to generalized badly to the remaining one-third of the uncommon words \cite{liu2021recent}. In this work, we use the same two third of the uncommon words from 16 dysarthric speakers, but all common and uncommon words from 13 control speakers as the training data (40-hours after silence stripping and about 190-hours after speed perturbation). Generalization to unseen words of E2E ASR model is out of scope of this paper. The evaluation data was collected from only the 16 dysarthric speakers and cover all 155 common words and the remaining one-third of the uncommon words (9-hours after silence stripping).} The source-target domain utterance length ratio between Librispeech and UASpeech is 12.3s:1.1s on average. A word grammar language model was used in decoding \cite{christensen2012comparative}.\\
\textbf{In the domain adaptation experiments}, all the source domain trained Conformer systems used 5000 Byte Pair Encoding (BPE) tokens selected from the 960-hour Librispeech transcripts as the decoder outputs. During domain adaptation, the normal speech pre-trained parameters in Conformer encoder and Transformer decoder were fine-tuned to dysarthric or elderly speech, while the other components of the systems were randomly reinitialized and retrained on the target domain data from scratch. In this process, 100 BPE tokens extracted from the DBank speech transcripts and character level tokens were used in fine-tuning the decoder module for the DBank and UASpeech data respectively.
%\footnote{Ablation studies suggest alternative cross-domain adaptation settings involving other combinations over the encoder and decoder modules: a) parameter re-estimation; b) fixing the pre-trained parameters; c) fine-tuning the pre-trained parameters; or d) using character based decoder outputs all led performance degradation.}.
\vspace{-0.2cm}
\subsection{Performance on DementiaBank}
\textbf{Baseline and manually designed system} The performance of the baseline system as described in Section 2.1 is shown in the first line (Sys. 1) in Tab. \ref{tab:dbank}. We increase the number of decoder Transformer blocks from 6 to 12 and reduce the convolution kernel size from 31 to 7. The resulting manually designed system outperformed the baseline (Sys. 2 vs. Sys. 1, Tab. \ref{tab:dbank}) by 
%significant
WER reduction of 2.75\% absolute on average, while the model parameters increased from 43.1M to 52.5M. \\
\textbf{Parameter and hyper-parameter adaptation} Among the results shown in Tab. \ref{tab:dbank}, several trends can be found. {\bf 1)} Using manually crafted hyper-parameters, the parameter domain adaptation of normal non-aged 960-hour Librispeech pre-trained Conformer systems produced 
% statistically significant 
WER reductions of 7.17\% absolute (22.34\% relative) on average (Sys. 6 vs. Sys. 2).
{\bf 2)} The hyper-parameter adaptation step consistently produced further WER reductions of 
%% 0.04\%-0.45\% 
up to 0.45\% on average on top of those obtained by parameter only fine-tuning (Sys. 8 vs Sys. 6) irrespective of the complexity penalty term $\eta$ setting in Eqn. (2). By further tuning the penalty $\eta$ during hyper-parameter adaptation, the most compact system achieves 0.24\% WER reductions with relatively 11.2\% smaller model parameters (Sys. 9 vs Sys. 6). {\bf 3)} Compared with hyper-parameter learning using only the in-domain DBank data (Sys. 3-5), the hyper-parameter domain adaptation step can automatically increase the model capacity to better leverage the knowledge distilled from normal non-aged 960-hour Librispeech pre-trained ASR systems (Sys. 7-9). 
\vspace{-0.2cm}
%As shown in Table \ref{tab:dbank}, irrespective of the architecture being baseline or manually designed, parameter domain adaptation of normal non-aged 300-hour Librispeech pre-trained Conformer systems (output token bpe2000) consistently produced statistically significant WER reductions of 7.5\%-8.8\% absolute (23.4\%-25.3\% relative) on average across both test sets over these without parameter adaptation (Sys. 6, 7 vs. Sys. 1, 2). 
%In addition, the use of penalized Gumbel-Softmax DARTS produced the most compact and best performed architecture by searching based on the manually designed system on DementiaBank data only, reaching architectural compression of 38.9\% by optimizing the feedforward and attention head dimensionality while incurring no performance degradation (Sys. 5 vs. Sys. 2). Finally, 
%However, we observe that the manually designed system with parameter adaptation achieves same level performance with the one adapted on both parameters and architecture regardless of the penalization (Sys. 7 vs Sys. 8-10). This is because the source and target domain have similar average utterance length in this task (5.3s vs 3.4s), and it is less necessary to vary the hyper-parameters such as convolution module that are designed to control the receptive field of local information.
\begin{table*}[tb]\caption{\label{tab:dbank}\setstretch{0.8}\fontsize{7.5bp}{7.5bp}\textnormal{{Performance (WER\%, \#Params) of Conformer systems with baseline, manually designed and automatically learnt hyper-parameter settings  trained on in-domain DBank data only (Sys. 1-5), and those cross domain parameter plus hyper-parameter adapted (Sys. 6-9). The manually designed Conformer (Sys. 2) serves as the start point of hyper-parameter in-domain learning and cross-domain adaptation performed over four groups of hyper-parameters inside each Conformer encoder block: a) the macron-feedforward and feedforward layer dimensionality (FD), where the dimensionality indices denote a choice from \{512, 1024, 2048, 3072\}; b) the number of attention heads (AH); c) the dimensionality of attention head (ADIM), where the indices denote a choice from \{16, 32, 64, 96\}; d) the convolution kernel size (CK), as well as three groups of hyper-parameters inside each Transformer decoder block (FD, AH and ADIM) with the same corresponding search space as in encoder. 
%The number of hyper-parameter selection weights are $S_{e} \times L_e + S_d \times L_d$ in total, where the $L_e=12$ and $L_d=12$ are the the number of blocks of encoder and decoder respectively, $S_e=4^2 \times 3 \times 4 \times 3 = 576$ and $S_d=4^2 \times 3 \times 4 = 192$ are the number of candidate architectures in the search scope inside each underlying encoder and decoder block.
$\eta$ is the penalty factor penalized DARTS of Eqn. (2).} "Inv" and "Par" refer to the WER for elderly participants and the clinical investigators respectively. 
$\dag$ denotes a statistically significant WER difference (MAPSSWE \cite{gillick1989some}, $\alpha$ = 0.05) is obtained over the baseline (Sys. 1).}}
\vspace{-0.6cm}
\scriptsize
\renewcommand\arraystretch{1.3}
\begin{center}
\scalebox{0.85}[0.85]{
\begin{tabular}{c|c|c|c|c|c|c|p{5mm}p{5mm}|p{5mm}p{5mm}|c|c}
\hline \hline
\multirow{2}{*}{Sys}           & \multirow{2}{*}{\begin{tabular}[c]{@{}c@{}}Parameter\\Adaptation\end{tabular}}   & \multicolumn{5}{c|}{Hyper-parameter Adaptation}            & \multicolumn{2}{c|}{Dev}        & \multicolumn{2}{c|}{Eval}       & \multirow{2}{*}{Avg.} & \multirow{2}{*}{\#Params} \\ \cline{3-7} \cline{8-11}
&                      &   \multicolumn{1}{c|}{Adaptation}      &   SearchMethod  & $\eta$     & \multicolumn{1}{c|}{Encoder Search Scope}                                    & Decoder Search Scope        & \multicolumn{1}{c|}{Inv} & Par  & \multicolumn{1}{c|}{Inv} & Par  &                      &                           \\ \hline \hline
1     & \multirow{5}{*}{$\times$}      & \multirow{5}{*}{$\times$}   & Baseline  & \multicolumn{3}{c|}{\multirow{2}{*}{$\times$}}        &             21.29            & 48.90 &                 19.20    & 37.35&         34.84         &       43.1M               \\ \cline{1-1} \cline{4-4}\cline{8-13}
2     &         &          & Manual &   \multicolumn{3}{c|}{}  & 19.53                     & 44.90 & 19.20                     & 34.66 & 32.09                 & 52.5M                     \\ \cline{1-1} \cline{4-7} \cline{8-13} 
3    &   &       &  \multirow{3}{*}{DARTS}       & 0 & \multirow{3}{*}{\begin{tabular}[c]{@{}c@{}}FD:\{0,1,2,3\}, AH:\{2,4,8\}\\ ADIM:\{0,1,2,3\}, CK:\{3,5,7\}\end{tabular}}         & \multicolumn{1}{r|}{\multirow{3}{*}{\begin{tabular}[c]{@{}c@{}}FD:\{0,1,2,3\}, AH:\{2,4,8\}\\ ADIM:\{0,1,2,3\}\end{tabular}}}  & 19.32&44.59&17.75&34.95& 31.89 &       53.2M            \\ \cline{1-1}  
4    &   &       &  & 0.003        & &  & 19.03&44.68&18.64&34.20&31.71&        39.3M           \\ \cline{1-1} 
5    &   &       &  & 0.03        & && 19.21&45.55&18.09&35.27&32.29&    22.6M \\ \hline \hline
6    &   \multirow{4}{*}{\begin{tabular}[c]{@{}c@{}}Librispeech\\$\rightarrow$\\Dbank\end{tabular}}   &  $\times$  &  Manual  &  \multicolumn{3}{c|}{$\times$} &15.27&35.11&17.20& 25.61 &24.92&    52.5M \\\cline{1-1} \cline{3-13}
7    &    &    \multirow{3}{*}{\begin{tabular}[c]{@{}c@{}}Librispeech\\$\rightarrow$\\Dbank\end{tabular}}    &  \multirow{3}{*}{DARTS}   & 0.        &  \multirow{3}{*}{\begin{tabular}[c]{@{}c@{}}FD:\{0,1,2,3\}, AH:\{2,4,8\}\\ ADIM:\{0,1,2,3\}, CK:\{3,5,7\}\end{tabular}}         & \multicolumn{1}{r|}{\multirow{3}{*}{\begin{tabular}[c]{@{}c@{}}FD:\{0,1,2,3\}, AH:\{2,4,8\}\\ ADIM:\{0,1,2,3\}\end{tabular}}}  & \textbf{14.88$\dag$}&35.47&15.76&25.74&24.88&    69.8M \\\cline{1-1} \cline{5-5} \cline{8-13}
8    &    &     &  & 0.003        & & & 15.03& \textbf{34.41$\dag$}&15.42&\textbf{25.48$\dag$}&\textbf{24.47$\dag$}&    65.7M \\\cline{1-1} \cline{5-5} \cline{8-13}
9    &     &       &  & 0.03        & & &15.20 &34.71&\textbf{14.76$\dag$}&25.80&24.68&    46.6M  \\
\hline \hline

\end{tabular}}
\end{center}
\vspace{-0.6cm}
\end{table*}

\begin{table*}[th]\caption{\label{tab:uaspeech}\setstretch{0.8}\fontsize{7.5bp}{7.5bp}\textnormal{Performance (WER\%, \#Params) of Conformer systems with baseline, manually designed hyper-parameter settings trained on in-domain UASpeech data only (Sys. 1-2), and those cross domain parameter plus hyper-parameter adapted (Sys. 3-6). $\dag$ and $\ddag$ denote a statistically significant WER difference is obtained over the baseline (Sys. 1) and the parameter-only fine-tuned system (Sys. 3) repectively. Other naming conventions following those of Tab. \ref{tab:dbank}.}}
\vspace{-0.6cm}
\scriptsize
\renewcommand\arraystretch{1.3}
\begin{center}
\scalebox{0.85}[0.85]{
\begin{tabular}{c|c|c|c|c|c|c|p{5mm}p{5mm}p{5mm}p{6mm}|c|c}
\hline \hline
\multirow{2}{*}{Sys}           & \multirow{2}{*}{\begin{tabular}[c]{@{}c@{}}Parameter\\Adaptation\end{tabular}}   & \multicolumn{5}{c|}{Hyper-parameter Adaptation}            &  \multirow{2}{*}{High}     & \multirow{2}{*}{Mid}  & \multirow{2}{*}{Low} & \multirow{2}{*}{\begin{tabular}[c]{@{}c@{}}Very\\Low\end{tabular}}  & \multirow{2}{*}{Avg.} & \multirow{2}{*}{\#Params} \\ \cline{3-7} 
&                      &   \multicolumn{1}{c|}{Adaptation}      &   SearchMethod  & $\eta$     & \multicolumn{1}{c|}{Encoder Search Scope}                                    & Decoder Search Scope        &  &   & &   &                      &                           \\ \hline \hline
1     & \multirow{2}{*}{$\times$}      & \multirow{2}{*}{$\times$}   & Baseline  & \multicolumn{3}{c|}{\multirow{2}{*}{$\times$}}        &   11.98 & 35.03 & 41.85 & 66.54&     35.71   &       43.1M               \\  \cline{4-4} \cline{1-1} \cline{8-13}
2     &      &  & Manual  & \multicolumn{3}{c|}{}        &             9.29            & 31.90 &      40.63    & 67.34 &         34.06         &       52.5M               \\ \hline \hline
3    &  \multirow{4}{*}{\begin{tabular}[c]{@{}c@{}}Librispeech\\$\rightarrow$\\UASpeech\end{tabular}}   & \multirow{1}{*}{$\times$}   & Manual  & \multicolumn{3}{c|}{\multirow{1}{*}{$\times$}}   &5.45& 23.21& \textbf{33.15$\dag$}&63.40&28.32&    52.5M \\\cline{1-1} \cline{3-13}
4    &    &    \multirow{3}{*}{\begin{tabular}[c]{@{}c@{}}Librispeech\\$\rightarrow$\\UASpeech\end{tabular}}    &  \multirow{3}{*}{DARTS}   & 0.        &  \multirow{3}{*}{\begin{tabular}[c]{@{}c@{}}FD:\{0,1,2,3\}, AH:\{2,4,8\}\\ ADIM:\{0,1,2,3\}, CK:\{3,5,7\}\end{tabular}}         & \multicolumn{1}{r|}{\multirow{3}{*}{\begin{tabular}[c]{@{}c@{}}FD:\{0,1,2,3\}, AH:\{2,4,8\}\\ ADIM:\{0,1,2,3\}\end{tabular}}}  & 5.22  & \textbf{21.35$\dag\ddag$} &33.37 & \textbf{62.06$\dag\ddag$} &\textbf{27.65$\dag\ddag$}&    47.8M \\\cline{1-1} \cline{5-5} \cline{8-13}
5    &    &     &  & 0.003      &  &  &5.38  & 22.49 &33.31 & 63.56 &28.23&    41.0M \\\cline{1-1} \cline{5-5} \cline{8-13}
6    &     &       &  & 0.03        & && \textbf{5.08$\dag$}&23.15& 34.21 &63.10&28.39&    29.8M  \\
\hline \hline
\end{tabular}}
\end{center}
\vspace{-1cm}
\end{table*}
\vspace{-0.3cm}
\begin{figure}[htb]
    \centering
    \includegraphics[width=0.45\textwidth]{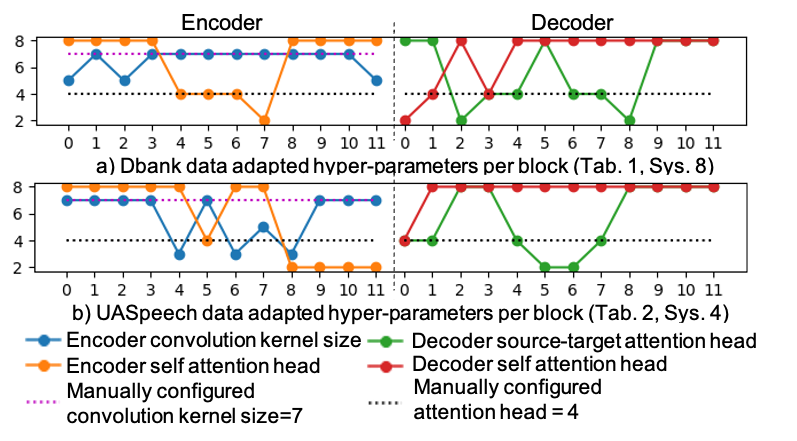}
    \vspace{-0.4cm}
    \caption{\rm \setstretch{0.8}\fontsize{7.5bp}{7.5bp}\textnormal{Hyper-parameters after cross-domain hyper-parameter adaptation from Librispeech to DBank (a) and from Librispeech to UASpeech (b). X axis is the encoder and decoder blocks where '0' refers to the very bottom layer close to input and '11' refers to the top layer close to CTC or output layer. Y axis represents two gropus of hyper-parameters including the number of attention heads and the convolution kernel size.}}\label{fig:hyper}
    \vspace{-0.6cm}
\end{figure}
\vspace{-0.2cm}
\begin{figure}[htb]
    \centering
    \includegraphics[width=0.45\textwidth]{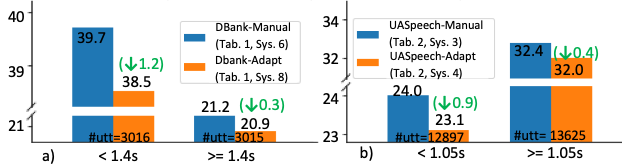}
    \vspace{-0.3cm}
    \caption{\rm \setstretch{0.8}\fontsize{7.5bp}{7.5bp}\textnormal{Conformer systems Performance (WER\%) on subsets of DBank (a) and UASpeech (b) test data of shorter and longer utterance lengths, with manually designed or cross domain adapted architecture hyper-parameters.}}\label{fig:werlen}
    \vspace{-0.6cm}
\end{figure}
\subsection{Performance on UASpeech}
\textbf{Baseline and manually designed system} is shown in Sys. 1 and 2 in Tab. \ref{tab:uaspeech} with the same hyper-parameter settings as DBank systems. The manually designed system outperformed the baseline (Sys. 2 vs. Sys. 1, Tab. \ref{tab:uaspeech}) by a 
%significant
WER reduction of 1.65\% absolute on average.\\
\textbf{Parameter and hyper-parameter adaptation} 
Several trends can be found in Tab. \ref{tab:uaspeech}. {\bf 1)} The parameter adaptation on the UASpeech manually designed system produces a WER reduction of 5.74\% absolute (16.85\% relative) on average (Sys. 3 vs. Sys. 2). Such relative WER reduction from parameter adaptation is smaller compared to the experiments on the DBank task (22.34\% relative, Sys. 6 vs. Sys. 2, Tab. \ref{tab:dbank}). This is due to the larger mismatch in terms of both utterance lengths and speech contents (sentences vs. single words) between the Librispeech and UASpeech data.  {\bf 2)}, The additional hyper-parameter adaptation step produced further statistically significant 
 consistent WER reductions of 
%% 0.09\%-0.67\% 
up to 0.67\%
on average of those obtained by parameter only fine-tuning (Sys. 4 vs Sys. 3).
%%with zero or small} complexity penalty term $\eta$ setting in Eqn. (2), \textcolor{red}{while large complexity penalty term (Sys. 6) leads to over-regularized small models, resulting in a mild increase of WER (0.07\%) over Sys. 3. 
The best-performed system achieves 1.34\% WER reductions on the most challenging "Very Low" intelligibility subsets, with relatively 9.0\% smaller model parameters (Sys. 4 vs Sys. 3).  {\bf 3)} Compared with the results on the DBank data of Tab. \ref{tab:dbank}, it is also found that the WER reductions from hyper-parameter domain adaptation are intuitively correlated with the relative utterance length ratio between the source and target domain data. This ratio is smaller between Librispeech and DBank (12.3s:3.4s) while larger between Librispeech and UASpeech (12.3s:1.1s). Such hypothesis is further analysed below in Section 4.3.
%% We present a detailed analysis of this hypothesis in Section 4.3.
\vspace{-0.2cm}
\subsection{Analysis of the Relation Between Hyper-parameter Adaptation and Source-target Utterance Length Ratio}
\vspace{-0.2cm}
In this section, the correlation between the performance improvements from Conformer hyper-parameter adaptation and the average utterance length ratio between the source and target domain is further analysed. This is conducted by evaluating the WER reductions on the subsets of DBank and UASPeech test data of shorter and longer utterance lengths. On both tasks the test data are divided using the respective median utterance length into "shorter" and "longer" subsets.    
As shown in Fig. \ref{fig:werlen}, on the shorter segments based subsets of DBank and UASpeech test data, hyper-parameter adaptation produces larger 
% and statistically significant 
WER reductions of 1.2\% and 0.9\% respectively over parameter-only fine-tuning. In contrast, smaller WER reductions of 0.3\%-0.4\% were obtained on the longer segments based subsets. These trends are also consistent with the corresponding larger changes of Conformer hyper-parameters on UASpeech data observed among more blocks when compared with those on DBank (overlapping with dotted lines of manual hyper-parameter settings indicate no change): a) the number of encoder attention heads that affect longer range contexts (Fig. \ref{fig:hyper}, (a) vs. (b) orange); and b) the kernel sizes of the convolution modules designed to control the span of local contexts (Fig. \ref{fig:hyper}, (a) vs. (b), blue). The performance of the best performing Conformer system using hyper-parameter domain adaptation is further contrasted with recently published results on the UASpeech and DBank task in Table \ref{tab:comparison}.
\begin{table}[htb]
\vspace{-0.3cm}
\caption{\label{tab:comparison}\setstretch{0.8}\fontsize{7.5bp}{7.5bp}\textnormal{A comparison of average WER between published systems and our hyper-parameter adapted system on DBank and UASpeech.}}
\vspace{-0.6cm}
\scriptsize
\renewcommand\arraystretch{1.3}
\begin{center}
\scalebox{0.85}[0.85]{
	\begin{tabular}{c|c}
	    \hline\hline
         DBank Systems & Avg. WER. \\
		\hline
		2021 Kaldi TDNN + Domain Adaptation \cite{ye2021development} & 32.33 \\
		2022 Conformer + Domain Adaptation \cite{wang2022conformer} & 25.60 \\
		\textbf{Conformer + Hyper-parameter Adaptation(Sys. 8, Tab. \ref{tab:dbank})} & \textbf{24.47} \\\hline \hline
		UASpeech Systems & Avg. WER. \\
		\hline
		2019 Kaldi TDNN + DA \cite{xiong2019phonetic} & 30.01 \\
		2020 DNN + DA \cite{geng2020investigation} & 28.73 \\
		\textbf{Conformer + Hyper-parameter Adaptation(Sys. 4, Tab. \ref{tab:uaspeech})} & \textbf{27.65} \\\hline \hline
  	\end{tabular}
	}
\vspace{-1cm}
\end{center}
\end{table}
\vspace{-0.1cm}
\section{Conclusions}
\vspace{-0.1cm}
This paper investigates hyper-parameter adaptation for Conformer ASR systems that are pre-trained on the Librispeech corpus before being domain adapted to the DementiaBank elderly and UASpeech dysarthric speech datasets. Experimental results suggest that conventional parametric domain adaptation from Librispeech data produced consistent WER reductions of up to 7.17\% and 5.74\% absolute (22.34\% and 16.85\% relative) for the DBank and UASpeech tasks respectively over the in-domain data trained Conformer models. Hyper-parameter adaptation produced further 
% statistically significant 
absolute WER reductions of 0.45\% and 0.67\% over parameter-only fine-tuning, and can better account for domain mismatch. An intuitive correlation is found between the performance improvements by hyper-parameter domain adaptation and the relative utterance length ratio between the source and target domain data.
\vspace{-0.2cm}
\section{Acknowledgements}
\vspace{-0.1cm}
\label{ssec:a}
This research is supported by Hong Kong RGC GRF grant No. 14200021, 14200218, 14200220, TRS T45-407/19N and Innovation \& Technology Fund grant No. ITS/254/19, ITS/218/21.

\bibliographystyle{IEEEtran}
\bibliography{mybib}

\end{document}